\documentclass[letterpaper]{article}
\usepackage{jheppub}
\usepackage{graphicx, caption, subcaption}  
\usepackage{dcolumn}   

\title{ Flavour issues in warped custodial models:  $B$ anomalies and rare $K$ decays} 
\author[a]{Giancarlo D'Ambrosio,}
\emailAdd{gdambros@na.infn.it}
\author[b]{Abhishek M. Iyer}
\emailAdd{iyera@na.infn.it}
\affiliation[a,b]{INFN-Sezione di Napoli, Via Cintia, 80126 Napoli, Italia}

\abstract{We explore the flavour structure of custodial Randall-Sundrum (RS) models in the context of  semi-leptonic decay of the $B$ mesons.  Anomalies in the $b\rightarrow s ll$ processes can be easily fit with partially composite second generation leptons and third generation quarks.
	Given the explanations of the $B$ anomalies, we obtain predictions for rare $K$- decays which are likely to be another candle for new physics (NP).
	 Two scenarios are considered: A)  The source of non-universality is the right handed muons (unorthodox case) B) Standard scenario, with anomalies explained primarily by non-universal couplings to the lepton doublets.
	The prediction for the rare $K$-decays are different according to the scenario, thereby serving as a useful discriminatory tool. We note that, in this setup $R(D^*)$ is at best consistent with the SM and increasing the compositeness of the $\tau_L$  generates a net contribution becoming below the SM expectation. 
Finally, we also comment on the implications of flavour violation in the lepton sector and present an explicit example with the implementation of bulk leptonic MFV which helps in alleviating the constraints. 
}

\newpage
\begin{document}
\maketitle
\flushbottom
\newpage


\section{Introduction}

Flavour physics, both in the lepton and hadron sector, offers an exciting avenue to  explore scales  even beyond the realm of the LHC. Processes like $\mu\rightarrow e\gamma$, $\tau\rightarrow \mu\gamma$ in leptonic sector and  $K_L\rightarrow \pi^0\nu\nu$, $K^+\rightarrow \pi^+\nu\nu$ ($s\rightarrow d$ transitions) in the hadronic sector are characterised by   small SM contributions thereby offering a  
 a lot of scope for the manifestation of NP. More recently, the LHCb has been involved in the measurement of the $b\rightarrow s ll$ flavour observables through the measurement  of the ratio \cite{Aaij:2014ora} 
\begin{eqnarray}
R_K=\left.\frac{\mathcal{B}(B^+\rightarrow K^+\mu^+\mu^-)}{\mathcal{B}(B^+\rightarrow K^+e^+e^-)}\right\vert_{q^2=1-6~GeV^2}=0.745^{+0.090}_{-0.074}~(stat)\pm 0.036~(syst)\nonumber\\
\end{eqnarray}
while the SM expectation is $R_K^{SM}=1.0003\pm0.0001$ \cite{Bobeth:2007dw,Bordone:2016gaq}. This implies a  deviation of $\sim 2.6~\sigma$ as a possible evidence of Lepton flavour universality (LFU) violation. 
The ratio, originally proposed in \cite{Hiller:2003js}, is a very clean test of the SM, as the hadronic uncertainties cancel.
This was further corroborated by the  measurement of the following ratio
\begin{eqnarray}
	R_{K^*}=\frac{\mathcal{B}(B^0\rightarrow K^{*0}\mu^+\mu^-)}{\mathcal{B}(B^0\rightarrow K^{*0}e^+e^-)}=\left\{ \quad
	\begin{aligned}
		\hspace{0cm} 0.660^{+0.110}_{-0.070}(stat) \pm 0.024(syst),\;\;\;0.045\leq q^2\leq 1.1 ~\text{GeV}^2 \\
		0.685^{+0.113}_{-0.069}(stat) \pm 0.047(syst),\;\;\;1.1\leq q^2\leq 6.0 \quad~\text{GeV}^2  \\
	\end{aligned}
	\right. \quad
\end{eqnarray}
The SM prediction in the corresponding $q^2$ bins are: $R^{SM}_{K^*}\simeq 0.93$ for low $q^2$ while $R^{SM}_{K^*}=1$ elsewhere. This corresponds to a $2.4\sigma$ deviation for low $q^2$ and $\sim 2.5~\sigma$ for medium $q^2$.
Further, the LHCb \cite{Aaij:2013qta,Aaij:2015oid} and the BELLE \cite{Wehle:2016yoi} collaboration have observed a deviation in the measurement of the angular observable $P'_5$ \cite{DescotesGenon:2012zf} in $B\rightarrow K^*\mu\mu$ decays. This stresses the possibility of lepton non-universality,  in the $\mu$ sector in particular \cite{Descotes-Genon:2013wba,Beaujean:2013soa,Alonso:2014csa,Hurth:2013ssa,Ghosh:2014awa,Hiller:2014yaa,Hurth:2014vma}. However, in the following we will not necessarily restrict ourselves to this possibility.
These deviations can be parametrized by the additional contributions to the following effective operators \cite{Buchalla:1995vs}:
\begin{eqnarray}
\mathcal{L}\supset \frac{V^*_{tb}V_{ts}G_F\alpha}{\sqrt{2}\pi}\sum_i C_i\mathcal{O}_i
\label{effective}
\end{eqnarray} 
where $C_i=C^{SM}_i+\Delta C_i$, with the NP contributions being included in $\Delta C_i$. The different fits signal towards additional contributions to one or a combinations of the following operators:
\begin{eqnarray}
\mathcal{O}_9&=&(\bar s_L\gamma^\mu b_L)(\bar l \gamma_\mu l)\;\;\;\;\;\;\;\;\;\mathcal{O'}_{9}=(\bar s_R\gamma^\mu b_R)(\bar l \gamma_\mu l)\nonumber\\
\mathcal{O}_{10}&=&(\bar s_L\gamma^\mu b_L)(\bar l  \gamma_\mu\gamma^5 l )\;\;\;\;\;\mathcal{O'}_{10}=(\bar s_R\gamma^\mu b_R)(\bar l \gamma_\mu\gamma^5 l)
\end{eqnarray}
There have been several analyses to determine the best fit values to the $\Delta C_i$:   1-D fits were performed and agreement with data can be  obtained if the NP satisfies one of the following hypotheses with the corresponding best fit points \cite{Capdevila:2017bsm}: 1) $\Delta C^\mu_9=-1.1$, 2) $\Delta C^\mu_9=-\Delta C^\mu_{10}=-0.61$ and 3) $\Delta C^{\mu}_9=-\Delta C^{'\mu}_{9}=-1.01$. In the 1-D hypotheses, the $\Delta C_i$ for the other operators in the effective theory are set to zero. In parallel, fits in the 2-D plane,  in   
$\Delta C^\mu_9-\Delta C^\mu_{10}$, $\Delta C^e_9-\Delta C^\mu_{9}$ and $\Delta C^{'\mu}_9-\Delta C^\mu_{9}$, respectively were performed in \cite{Altmannshofer:2017yso}. 
Furthermore, it is also possible to obtain a fit to the data in the 6-D parameter space  and obtaining the following best fit points \cite{Capdevila:2017bsm}:
\begin{equation}
\Delta C_{7}=0.017\;\;\;\Delta C^\mu_{9}=-1.12\;\;\;\Delta C^\mu_{10}=0.33\;\;\;\Delta C'_{7}=0.59\;\;\;\Delta C^{'\mu}_{9}=0.59\;\;\;\Delta C^{'\mu}_{10}=0.07
\end{equation} 
A global four operator Wilson co-efficient fit has been performed using the $R_K$ data in \cite{Hurth:2016fbr} and assuming NP contributions to both the electrons and the muons. These were shown to relax the stringent criteria on Wilson co-efficients in the 1-D and 2-D scenarios.
 Furthermore, if one assumes  $C_{LL}=C_{RL}=C_{RR}=0$ \footnote{These correspond to the Wilson coefficients in the chiral representation and are related to $C_{9,10}$ and $C'_{9,10}$ },a  large electron contribution to  $C_{LR}$ is necessary to satisfy the anomalies \cite{Hiller:2014ula,Hiller:2014yaa}. This is   is an explicit example where NP couples more to the electrons than the muons. A model independent analysis investigating the role of different operators in the possible explanations was considered in \cite{Bardhan:2017xcc,Ghosh:2017ber,Bardhan:2016uhr}.

There have been several proposed extensions to explain this anomaly, typically ascribing to one or more of the hypothesis above: models with leptoquarks \cite{Calibbi:2015kma,Alonso:2015sja,Hiller:2014yaa,Gripaios:2014tna,Sahoo:2015wya,Varzielas:2015iva,Becirevic:2015asa,Fajfer:2015ycq,Becirevic:2016oho,Calibbi:2017qbu,Crivellin:2017dsk, Barbieri:2017tuq}, scenarios with an additional $Z'$ \cite{Gauld:2013qba,Glashow:2014iga,Bhattacharya:2014wla,Crivellin:2015mga,Crivellin:2015lwa,Sierra:2015fma,Crivellin:2015era,Celis:2015ara,Belanger:2015nma,Gripaios:2015gra,Allanach:2015gkd,Fuyuto:2015gmk,Chiang:2016qov,Boucenna:2016wpr,Boucenna:2016qad,Celis:2016ayl,Altmannshofer:2016jzy,Crivellin:2016ejn,GarciaGarcia:2016nvr,Becirevic:2016zri,Bhattacharya:2016mcc,Bhatia:2017tgo,Cline:2017lvv}, extra-dimensional models with soft wall \cite{Megias:2016bde,Megias:2017ove,Megias:2017vdg}, partial compositeness \cite{Carmona:2017fsn,Sannino:2017utc}
 and RPV SUSY \cite{Bhattacharya:2016mcc,Das:2017kfo,Altmannshofer:2017poe}. 
A combination of model dependent and independent analysis was performed in \cite{Alok:2017sui}.

Further hints of lepton non-universality were also observed by BaBar \cite{Lees:2012xj,Lees:2013uzd}, LHCB \cite{Aaij:2015yra} and BELLE \cite{Hirose:2016wfn} in charged current transitions ($b\rightarrow c$) with the measurement of the following observables:
\begin{eqnarray}
R(D)&=&\frac{\mathcal{B}(\bar B\rightarrow D^+\tau^-\bar{\nu_\tau})}{\mathcal{B}(\bar B\rightarrow D^+l^-\bar{\nu_l})}=0.440\pm 0.058\pm 0.042\nonumber\\
R(D^*)&=&\frac{\mathcal{B}(\bar B\rightarrow D^{*+}\tau^-\bar{\nu_\tau})}{\mathcal{B}(\bar B\rightarrow D^{*+}l^-\bar{\nu_l})}=0.332\pm 0.024\pm 0.018\\
\textbf{where}~l=e,\mu \nonumber
\label{RD}
\end{eqnarray}

The SM expectation for these observables are $R(D)=0.297\pm0.017$ and $R(D^*)=0.252\pm0.003$ \cite{Fajfer:2012vx,Sakaki:2012ft} thereby indicating a 2.0 and $2.7\sigma$ deviation respectively from the experimental measurements. A unified description of both the $R(K)$ and $R(D^*)$ puzzles has been investigated extensively. For instance, in the prescription of \cite{Glashow:2014iga} where the NP couples only to the third generation fermion, a complete gauge invariant description of such operators will also result in charged current decays \cite{Fajfer:2012jt,Alonso:2014csa,Buras:2014fpa,Bhattacharya:2014wla,Calibbi:2015kma}.
However, as argued in \cite{Feruglio:2016gvd,Feruglio:2017rjo}, inclusion of quantum effects by means of RGE running from the NP scale $\Lambda$ to $\sim 1$ GeV, will also possibly introduce lepton flavour violating  (LFV) effects. As a result, explanation of $R(D^*)$ in this particular framework is in tension with the low energy LFV data from $\tau$ decays.
Other studies in this direction include: a comprehensive model analysis of different scenarios offering a simultaneous explanation was considered in \cite{Bhattacharya:2016mcc}, models with $U(2)^5$ symmetry groups and leptoquarks  \cite{Barbieri:2015yvd,Barbieri:2016las,Bordone:2017anc}, frameworks with $SU(2)_L$ triplet massive vector bosons \cite{Greljo:2015mma}, leptoquark scenarios \cite{Crivellin:2017zlb,Calibbi:2017qbu}. An interplay of the scales admissible by the explanation of the anomalies and collider implications was studied in \cite{Allanach:2017bta}.

In this paper we consider a model with a single warped extra-dimension compactified on a $S^1/Z_2$ manifold with the following line element \cite{Randall:1999ee}:
\begin{equation}
ds^2=e^{-2A(y)}\eta_{\mu\nu}dx^\mu dx^\nu-dy^2
\end{equation}
where $A(y)=k|y|$, $k\sim \frac{M_{Pl}}{4\pi}$ and $0\leq y \leq \pi R$. The coordinates $y=0,\pi R$ represent the location of the $(UV,IR)$ brane respectively. The effective UV cut-off on the IR brane ($M_{IR}$) is related to the one on the UV brane ($k$) as
\begin{equation}
M_{IR}=e^{-kR\pi}k\sim \mathcal{O}(TeV)~\text{for}~kR\sim 11
\end{equation}
This exponential warping of scales in an $AdS$ background is the celebrated solution to the hierarchy problem.
We consider a generalization of the original RS setup with a bulk custodial symmetry \cite{Agashe:2003zs}.
This setup is characterized by additional heavy gauge bosons in addition to the KK states of the SM $W,Z$ and consequently leads to a distinct phenomenology, in the flavour sector in particular. A detailed analysis of different flavour transitions in this setup was considered in \cite{Blanke:2008zb,Blanke:2008yr,Bauer:2009cf,Casagrande:2010si} and will form the basis of this analysis. We revisit this setup exploring the parameter space admitted by the current anomalies and offer predictions for the $K\rightarrow\pi\nu\nu$  decays, in the $s\rightarrow d$ sector. The SM expectation for the $K^+\rightarrow\pi^+\nu\bar{\nu}$ and $K_L\rightarrow\pi^0\nu\bar{\nu}$ is \cite{Buras:2006gb,Brod:2010hi,Buras:2015qea}:
\begin{equation}
\mathcal{B}(K^+\rightarrow\pi^+\nu\bar{\nu})=8.3\pm 0.3\pm 0.3 \times 10^{-11}\;\;\;\;\mathcal{B}(K_L\rightarrow\pi^{0}\nu\bar{\nu})=2.9\pm 0.2\pm 0.0 \times 10^{-11}
\end{equation}
where the first error is due to the uncertainty in the  $V_{CKM}$ parameters  while the second one corresponds to the remaining theoretical uncertainties.
The  current experimental bound is \cite{Patrignani:2016xqp}
\begin{equation}
\mathcal{B}(K^+\rightarrow\pi^+\nu\bar{\nu})=17.3^{+11.5}_{-10.5} \times 10^{-11}\;\;\;\;\mathcal{B}(K_L\rightarrow\pi^{0}\nu\bar{\nu})\leq 2.6\times 10^{-8}~~~~(90\%~\text{C.L.})
\end{equation}
These measurements are likely to be significantly improved in the future: The NA62 experiment at CERN \cite{Romano:2014xda,Rinella:2014wfa} is aiming to collect 20 SM events in 2018. 
while a  $5\%$ accuracy is likely to be achieved with  more time. 
Regarding $K_L \to \pi^0 \nu \overline{\nu}$, the KOTO experiment at J-PARC  aims  at 
measuring  $\mathcal{B}(K_L \to \pi^0 \nu \overline{\nu})$ around the SM sensitivity in the first instance~\cite{Komatsubara:2012pn,Shiomi:2014sfa}.
Moreover, the KOTO-step2 experiment will aim at 100 events for the SM branching ratio. This implies
a precision of $10\%$ of this measurement \cite{KOTO}.  Some NP scenarios where these processes were considered include leptoquarks \cite{Bobeth:2017ecx, Dorsner:2017ufx}, MSSM \cite{Crivellin:2017gks}, additional  gauge bosons \cite{Bordone:2017lsy} \textit{etc}.

Similar to \cite{Megias:2016bde,Megias:2017ove,Megias:2017vdg}, the anomalies in the $b\rightarrow s$ transitions can be achieved by a slightly composite third generation quark doublet.  The custodial protection prevents large contribution to $Z\rightarrow b_L\bar b_L$. We demonstrate fits with two  scenarios:\\ Scenario \textbf{A}. Right handed leptons are  likely to be more composite than the left handed leptons, in particular for the muon and tau. In this case non-universality exists in the right-sector while the coupling of the doublets to  NP is universal. A similar case was considered in \cite{Carmona:2017fsn},\\ Scenario \textbf{B}. Left handed lepton sector is more composite than the right handed leptons. \\
For  the former case, the NP contributes dominantly to $\Delta C^\mu_{9,10}$ and  with a smaller contribution to $\Delta C^e_{9,10}$. Therefore using  $4$-$D$ fit in \cite{Hurth:2016fbr}, we can obtain consistency with the data. It is to be noted that even though the first generation leptons are completely elementary, the new physics contribution to $\Delta C^e_{9,10}$ is non-zero. This can be attributed to the choice of wave function of the lepton doublets that characterizes a given scenario.
 The primed operators do not contribute as we assume universality in the bulk wave-functions of the right handed quarks.

The prediction for  rare  $K$ decays would be different in both the scenarios,  making it a useful discriminant.
Another interesting features of this scenario would be to check the compositeness of the $\tau$ lepton.
The net contribution to  $R(D), R(D^*)$ is consistent with the SM for the parameter space which fits the $\tau$ mass and reduces as the compositeness of the $\tau$ increases. This is mainly due to the large $W^{(0)}$-$W^{(1)}$ mixing which is proportional to the volume factor $\sqrt{2kR\pi}$.
  This makes it a predictive framework and a more accurate determination of $R(D), R(D^*)$ will help to shed more light on the underlying geometry involved.  
 For both  scenarios we choose parameters such that the  first two generation quarks couple universally to the NP gauge bosons. This results in an accidental $U(2)$ symmetry which are essential to alleviate constraints from $\Delta F=2$ FCNC processes \cite{Barbieri:2012tu,Panico:2016ull,Redi:2012uj,Barbieri:2012uh}.
 However, we perform further checks to explicitly determine the range of $c_{Q_3}$ consistent with these constraints.
 
  Non-universality in the lepton sector may also be a harbinger for dangerous LFV effects.
   Mixing of leptons with the KK states is known to give rise  to large contributions to FCNC like $\mu\rightarrow e\gamma$ \cite{Agashe:2006iy}. However it was shown that imposition of bulk MFV ansatz can alleviate these constraints \cite{Perez:2008ee,Chen:2008qg}. We demonstrate that the first scenario can easily accommodate the MFV ansatz albeit with a mild  tuning ($\sim 0.03$) in the muon Yukawa, $Y_\mu$ while the other charged lepton Yukawa coupling are chosen $\mathcal{O}(0.3)$. Along the way, we also present examples with fits to the neutrino oscillation data.  

The paper is organized as follows:  In Section \ref{model} we give a brief description of the custodial RS model and identify the parameter space consistent with the fermion mass fits. In Section \ref{deltaf} we limit the parameter space of third generation quarks to be consistent with $\Delta F=2$ processes.
In Section \ref{btos} we compute the fits for the anomalies in $b\rightarrow sll$ processes for two different 1-D hypotheses. In Section \ref{kaon} we consider the rare kaon decays and demonstrate how it can be utilised to possibly distinguish between the two scenarios used to fit the $b\rightarrow sll$ anomalies. In Section \ref{RDstar} we argue why this setup is best consistent with the SM when $R(D^*)$ is taken into account.
In Section \ref{MFV} we give an explicit example with MFV implemented in the lepton sector, in particular for the first scenario. In Appendix \ref{rsintro} we outline the structure the flavour violating couplings. 

\section{The Model}
\label{model}
The custodial RS model
 is characterized with an enlarged bulk gauge symmetry: $SU(2)_L\times SU(2)_R\times U(1)_{B-L}$. The Higgs doublet ($\phi$) is localized on the IR brane and  is promoted to a bi-doublet under  $SU(2)_L\times SU(2)_R$ as \cite{Agashe:2003zs,Blanke:2008yr,Blanke:2008zb}:
\begin{equation}
\Sigma=(\epsilon \phi^*,\phi)=\begin{bmatrix}
\phi^{*0}&\phi^+\\
-\phi^-&\phi^{0}
\end{bmatrix}
\end{equation}
where both $\phi$ and $\epsilon\phi^*$ are doublets of $SU(2)_L$.
The scalar Lagrangian on the IR brane is given as:
\begin{equation}
\mathcal{L}_{Higgs}=\int dy\delta(y-\pi R)\sqrt{-g}\left[Tr(D_M \Sigma)^\dagger D_M \Sigma-\mu^2Tr\Sigma^\dagger\Sigma+\lambda (Tr\Sigma^\dagger\Sigma)^2\right]
\end{equation}
where $\sqrt{-g}|_{y=\pi R}=e^{-4kr\pi}$. Re-defining the Higgs field as $\Sigma\rightarrow e^{kR\pi} \Sigma$ leads to the canonically normalized scalar Lagrangian.

\begin{figure}[htb!]
	\begin{center}
		\begin{tabular}{cc}
			\includegraphics[width=7.2cm]{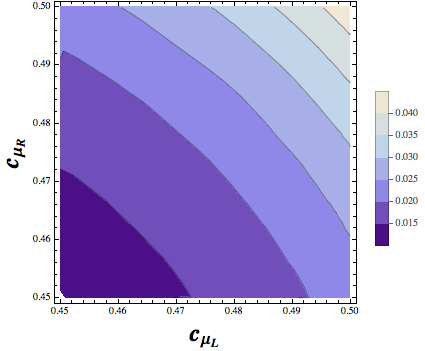}&\includegraphics[width=7.2cm]{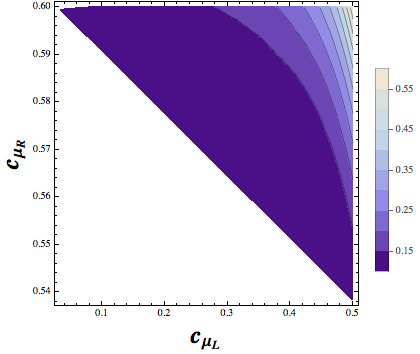}
		\end{tabular}
	\end{center}
	\caption{ Contours of fits muon Yukawa in the $c_{\mu_L}-c_{\mu_R}$ plane for Scenario A (left) and Scenario B(right). }
	\protect\label{comparison}
\end{figure}

The $\phi^{(0)}$ component develops a vev on the IR brane resulting in the symmetry breaking pattern $SU(2)_L\times SU(2)_R\rightarrow SU(2)_V$ and is responsible for the protection of the $T$ parameter.
On the UV brane, the bulk custodial symmetry is broken by orbifolding with the following choice of the  boundary conditions for the bulk gauge fields:
\begin{eqnarray}
W^a_{L\mu}(+,+)\;\;\;\;\;B_\mu(+,+)\;\;\;\;W^a_{R\mu}(-,+)\;\;\;\;Z_{X,\mu}(-,+)
\end{eqnarray}
As a result, the  residual symmetry on the UV brane is $SU(2)_L\times SU(2)_R\times U(1)_X\rightarrow SU(2)_L\times U(1)_Y$. In the light of the breaking on the UV and the IR brane, the effective low energy theory is $U(1)_Q$ symmetric. A discussion of the solutions for the bulk profiles of the gauge fields is given in Appendix \ref{rsintro}.\\
\textbf{Fermions:} The fermions in the theory, like the gauge bosons, are also bulk fields. The quarks and lepton doublets are embedded in a bi-doublet representation of $SU(2)_L\times SU(2)_R$ as
\begin{equation}
\zeta\equiv\begin{bmatrix}
\xi^u_L(-,+)&v^u(+,+)\\
\chi^d_L(-+)&v^d(+,+)
\end{bmatrix}
\end{equation} 
where $v=Q,L$ denotes a bulk  field whose zero mode corresponds to the SM quark(lepton) doublets. 
The superscripts $^{u(d)}$ are used to denote the $T_3=\frac{1}{2}(-\frac{1}{2})$ components of the doublet fields.
$\xi(\chi)$ are exotic fermions with $Q=5/3(2/3)$ which do not have a zero mode on account of the choice of boundary conditions.
The  singlets (right handed neutrinos $N$ and the up quarks) transform as $(1,1)$ while the charged lepton and down type singlets are embedded in a $\Psi\equiv (1,3)$ multiplet of bulk custodial symmetry. This ensures a custodial protection of $Z\rightarrow ff$ coupling for $f=b,l$ and aids in the realization of composite leptons which will be necessary for the explanation of lepton flavour universality violations in $B$ decays to be discussed later. Details of the solutions to the bulk gauge and fermion profiles are discussed in Appendix \ref{rsintro}.\\
\textbf{Yukawa Interactions:}
The SM fermions couple to the Higgs through brane localized Yukawa interactions of the form:
\begin{eqnarray}
\mathcal{L}_{Y}=\frac{Y_d^{'}}{k}\bar \zeta_Q \Sigma\Psi_d +\frac{Y_u^{'}}{k}\bar \zeta_Q \Sigma u_R+\frac{Y_N^{'}}{k}\bar \zeta_L \Sigma N+\frac{Y_E^{'}}{k}\bar \zeta_L \Sigma \Psi_e
\end{eqnarray}
$Y'$ are dimensionless Yukawa parameters and we have suppressed the flavour indices.
After EWSB, the effective  four dimensional Yukawa coupling is given as \cite{Gherghetta:2000qt}:
\begin{equation}
Y^{(4)}=2Y^{'}f_i(c_L)f_j(c_R)\;\;\;\;\text{where}~f(c)=\sqrt{\frac{0.5-c}{e^{(1-2c)kR\pi}-1}}e^{(0.5-c)kR\pi}
\end{equation}
Furthermore,  defining $Y^{(5)}=2Y^{'}$ as the  dimensionless $\mathcal{O}(1)$ Yukawa couplings, the effective Yukawa interaction is now gives as:\footnote{Some literature also follows the convention where $Y'$ is chosen as the $\mathcal{O}$ parameter to begin with and differs from the  $Y^{(5)}$ here by a factor of 2. }. 
\begin{equation}
Y^{(4)}=Y^{(5)}f(c_L)f(c_R)
\end{equation}
$c$ is the bulk mass parameter for the bulk fermion fields and is typically chosen between $-1.5<c<1.5$, where $c>0.5(<0.5)$ corresponds to fields localized close to UV (IR) brane. \\
To determine the extent to which non-universality can be accommodated in the model, the admissible range of $c$ parameters which fits fermion mass and mixing data need to be determined. They can be estimated by adapting the methodology in \cite{Iyer:2012db} by defining the following $\chi^2$ function:
\begin{equation}
\chi^2=\sum_{i}\frac{\left(\mathcal{O}^{exp}_i-\mathcal{O}^{theory}_i(c_j)\right)^2}{\sigma_i^2}
\label{chisq}
\end{equation}
where $\mathcal{O}^{exp}_i$ of the $i^{th}$ observable and $\mathcal{O}^{theory}_i(c_j)$ is the corresponding theoretical value as a function of $c_j$.
The variables in the hadronic and the leptonic sector are:
\begin{eqnarray}
\mathcal{O}^{hadronic}_i(c^h_j)&\equiv&\left[m_{u,c,t},m_{d,s,b},V_{CKM}\right]\nonumber\\
\mathcal{O}^{leptonic}_i(c^l_j)&\equiv&\left[m_{e,\mu,\tau},\Delta m^2_{sol},\Delta m^2_{atm},V_{PMNS}\right]
\end{eqnarray}
and both are fit separately \text{i.e.} the $\chi^2$ in Eq. \ref{chisq} is minimized separately for hadron and lepton sector. The scanning ranges are motivated by the constraints from flavour processes and the solutions to the LFU violations and are discussed below:\\
\textbf{Hadron mass fits:} The $c$ parameters for all the fields, with the exception of $c_{Q_3}$ and $c_{t_R}$  are chosen to be larger than $0.5$. This ensures universality in the right handed quark currents and a $U(2)$ symmetry among the first two generation doublets. We choose $c_{Q_3}\in[0.35,0.5]$ and $c_{t_R}\in[0,0.5]$. The relatively larger compositeness in the top singlet fields is to ensure a fit to the top mass without the requirement of large Yukawa couplings. The $\mathcal{O}(1)$ Yukawa couplings, $Y^{5}$, are chosen in the range $[0.1,5]$.\\
\textbf{Lepton mass fits:} For the lepton sector we consider two scenarios for the charged leptons which are motivated by the solutions to the flavour anomalies to be discussed later:\\
\textit{Scenario A:} The doublets have a universal bulk wave-function, with the $c$ parameters scanned in the range $0.45<c_L<0.55$. For the charged lepton singlets we choose the following ranges:
\begin{equation}
c_{e_R}\in[0.6,0.8]\;\;\;c_{\mu_R}\in[0.44,0.55]\;\;\;c_{\tau_R}\in[0.3,0.5]
\end{equation}
The compositeness of the right handed muon is to ensure fits consistent with data for LFU violations. This however results in minor fine tuning in the choice of the $Y^5_{\mu}$ as shown in the left panel of Fig. \ref{comparison}. It gives contours in the $c_{\mu_L}-c_{\mu_R}$ plane and the corresponding  $Y^5_{\mu}$, implying $\mathcal{O}(1)$ parameters $\sim 0.04$ to fit the muon mass.\\

\textit{Scenario B:} In this case, we consider non- universal doublet wave functions with a fair degree of compositeness for the muon and the tau doublets. We choose $c_{\mu_L}\in[0,0.5]$, $c_{\tau_L}\in[0.2,0.5]$ and $c_{e_L}\in[0.55,0.8]$. The $c$ values for singlets  are chosen larger than $0.50$. The right plot of Fig. \ref{comparison} gives the  gives contours in the $c_{\mu_L}-c_{\mu_R}$ plane with corresponding  $Y^5_{\mu}$:
it admits relatively less fine-tuned values in comparison to Scenario A.\\

\section{Constraints from different flavour observables}
\label{deltaf}
\begin{figure}[htb!]
	\begin{center}
		\begin{tabular}{cc}
			\includegraphics[width=7.2cm]{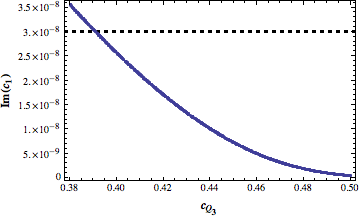}&\includegraphics[width=7.2cm]{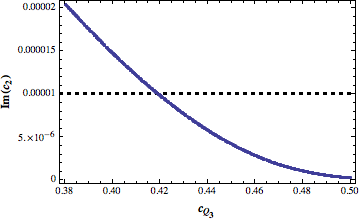}
		\end{tabular}
	\end{center}
	\caption{The Wilson co-efficient $c_i$ as a function of $c_{Q_3}$ for $s\rightarrow d$ and $b\rightarrow d$ transitions. }
	\protect\label{comparison1}
\end{figure}

An artefact of a model with a warped geometry is that all the gauge KK states are localized near the IR brane. 
Furthermore, as seen in Section \ref{model}, the localization of the fermionic generations at different points in the bulk results in varying degree of overlap with these heavy spin-1 resonances. Consequently, this gives rise to tree-level contributions to FCNC, in both the quark and the lepton sector. A discussion to this effect is presented in Appendix \ref{rsintro}.
 For the quark case, the non-universality exists between the third and the first two generations and its impact on some strongly constrained flavour observables will be discussed in this section.\\
\textbf{$\Delta F =2$ processes}:
While the main contribution to $\Delta F=2$ processes is due to the exchange of KK gluons, the contribution of the other gauge KK states cannot be ignored. 
In a generic RS framework,  the Hamiltonian  contributing to $\Delta F =2$ due to the exchange of first KK state is given as\cite{Blanke:2008yr,Blanke:2008zb}
\begin{eqnarray}
\mathcal{H}_{\Delta F=2}&=&\frac{1}{M^2_{KK}}\left[(a^{ij}_L)^2\left(\bar i_L\gamma_\mu t^a j_L\right)\left(\bar i_L\gamma^\mu t^a j_L\right) +(a^{ij}_R)^2\left(\bar i_R\gamma_\mu t^a j_R\right)\left(\bar i_R\gamma^\mu t^a j_R\right) \right.\nonumber\\&+&\left.a^{ij}_R a^{ij}_L\left(\bar i_R\gamma_\mu t^a j_R\right)\left(\bar i_L\gamma^\mu t^a j_L\right)  \right]
\end{eqnarray}
The WC for the flavour violating operators ($i\neq j$) are proportional to 
\begin{eqnarray}
a^{{12}}=\tilde g\left(D^*_{21}D_{22}(I(2)-I(1))+D^*_{31}D_{32}(I(3)-I(1))\right)\nonumber\\
a^{{23}}=\tilde g\left(D^*_{12}D_{13}(I(1)-I(2))+D^*_{32}D_{33}(I(3)-I(2))\right)\nonumber\\
a^{{13}}=\tilde g\left(D^*_{21}D_{23}(I(2)-I(1))+D^*_{31}D_{33}(I(3)-I(1))\right)
\end{eqnarray}
where $\tilde g$ is a generic parameter to denote the  coupling strength of gauge KK boson to a pair of fermions.
Here we choose $D=V_{CKM}$. $I(j),~j=1,2,3$ gives the overlap of the KK gauge boson wave function with that of the zero mode fermions and is defined in Eq \ref{couplingkkf}. 
In the scenario where $I(1)=I(2)<I(3)$, the contribution to the different $i$-$j$ transitions is simply:
\begin{eqnarray}
a^{{sd}}\propto V^*_{td}V_{ts}(I(3)-I(1))\nonumber\\
a^{{bd}}\propto V_{tb}V^*_{td}(I(3)-I(1))\nonumber\\
a^{{bs}}\propto V_{tb}V^*_{ts}(I(3)-I(1))
\end{eqnarray}
Further, in accordance with the parameter space scanned to fit the quark masses, the down quark singlets have universal coupling to all the  gauge KK bosons. As a result, $a^{ij}_R=0$. This implies that the operator  structure is exactly similar to that in the SM: \textit{viz.} $(V-A)(V-A)$. To determine the extent of the allowed contribution to the co-efficients $a^{ij}$ we consider the following parametrization of the  effective Lagrangian \cite{Isidori:2010kg,Isidori:2013ez,Gori:2016lga}:
\begin{equation}
\mathcal{L}_{eff}=\mathcal{L}_{SM}+\frac{c_1}{\Lambda^2}(\bar s_L\gamma^\mu d_L)^2+\frac{c_2}{\Lambda^2}(\bar b_L\gamma^\mu d_L)^2+\frac{c_3}{\Lambda^2}(\bar b_L\gamma^\mu s_L)^2
\end{equation}
Table \ref{constraints} gives the upper bounds on $c_i$ for $\Lambda=M_{KK}= 3$ TeV \cite{Isidori:2010kg,Isidori:2013ez,Gori:2016lga}
Fig. \ref{comparison1} gives as computation of $c_i$ as a function of $c_{Q_3}$. Evidently, as $c_{Q_3}$ increases to $0.5$, the breaking of the $U(3)$ to $U(2)$ symmetry is increasingly softer, thereby reducing the corresponding contributions.
 We find that $c_{Q_3}\gtrapprox 0.4$ is roughly preferred by the upper bound on the co-efficients in Table \ref{constraints}.\\ 
\begin{table}[htb!]
	\begin{center}
		\begin{tabular}{ |c|c | c |  }
			\hline
			Process	&$Re(c_i)$&$Im(c_i)$\\
			\hline			
			$(\bar s_L\gamma^\mu d_L)^2$&$8.1\times 10^{-6}$ & $3.0\times 10^{-8}$  \\
			$(\bar b_L\gamma^\mu d_L)^2$	&$2.0\times 10^{-5}$ &$9.9\times 10^{-6}$ \\
			$(\bar b_L\gamma^\mu s_L)^2$	&$4.5\times 10^{-4}$&$1.5\times 10^{-6}$\\
			\hline  
		\end{tabular}
	\end{center}
	\caption{Upper bounds on the Wilson coefficients of operators contributing to $\Delta F=2$ processes corresponding to $M_{KK}=3$ TeV.} 
	\label{constraints}
\end{table}

\textbf{$b\rightarrow s\gamma$:}
We now discuss the effects of composite third generation quarks on loop induced processes. The most dominant contribution would be to $b\rightarrow s\gamma$.
The relevant operators are given as
\begin{equation}
\mathcal{O}_7\equiv \bar b_R\sigma^{\mu\nu}s_LF^{\mu\nu}\;\;\;\;\;\;\;\mathcal{O'}_7\equiv \bar b_L\sigma^{\mu\nu}s_RF^{\mu\nu}
\end{equation}
Consider the contribution to the Wilson co-efficient of $\mathcal{O'}_7$ which is suppressed in the SM.
In RS it receives corrections due to A) KK gauge bosons and  charged  fermions in the internal lines and b) Higgs and  KK fermions in the internal lines. We consider each of them separately.\\
A) For the scenario with KK gauge bosons, the Wilson coefficient $C'_7$, for a generic $i\rightarrow j\gamma$ process is given as \cite{Agashe:2004cp}
\begin{equation}
(C'_7)_{ij}\propto F^i_LY_{ij}F^j_R
\end{equation}
where the factor $Y$ is due to mass insertion in the internal  fermion line and $F$ is $3\times3$ matrix of zero mode wavefuction in flavour space defined as:
\begin{equation}
F=\begin{bmatrix}
f(c_1)&0&0\\
0&f(c_2)&0\\
0&0&f(c_3)
\end{bmatrix}
\end{equation} The flavour structure in this case is exactly aligned with the fermion mass matrix and hence the contribution is negligible.\\

\begin{figure}[htb!]
	\begin{center}
		\begin{tabular}{c}
			\includegraphics[width=7.2cm]{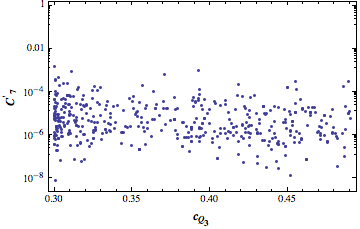}
		\end{tabular}
	\end{center}
	\caption{ $C'_7$ as a function of $c_{Q_3}$ }
	\protect\label{Cprime}
\end{figure}

B) In this case, there are three Yukawa insertions: two on the Higgs-fermion-KK fermion vertex and one on the internal KK fermion line. The flavour structure in this case is given as \cite{Agashe:2004cp}:
\begin{eqnarray}
C'_7\sim \frac{m_b}{v^2M^2_{KK}}\left[V^\dagger_{CKM}M^{diag}_uU_R^\dagger F^{-2}_UU_RM^{diag}_uU_L^\dagger F_Q^{-2}D_L+m_sD_R^\dagger F^{-2}_DD_RM^{diag}_dD_L^\dagger F_Q^{-2}D_L  \right]
\end{eqnarray}
Bounds exist on the values of $C'_7$ which can be extracted from the operator structure defined below:
\begin{equation}
\mathcal{H}_{eff}=\frac{c'_7}{\Lambda^2}~\mathcal{O'}_7\;\;\;\;\text{where}~\mathcal{O'}_7=\frac{m_b}{e}(\bar s\sigma_{\mu\nu}P_L b)F^{\mu\nu}
\end{equation}
The upper bound on $c_i$ for $\Lambda=1$ TeV is $3.6\times 10^{-4}$ \cite{Altmannshofer:2012az}. Fig. \ref{Cprime} gives the co-efficient as a function of $c_{Q_3}$. The elements of the rotation matrix for the up and down sector are obtained from the scan using $\chi^2$ minimization. We find that the co-efficient admits a rather democratic distribution and independent of the value of $c_{Q_3}$. This can be attributed to the pattern of $\mathcal{O}(1)$ which were typically chosen between 0.1 and 5 in the fit. As a result we conclude, that the dominant constraint to the range of $c_{Q_3}$ is due to the $\Delta F=2$ processes.

\section{$b\rightarrow sll$ processes}
\label{btos}
The explanations to the observed deviations in the measurements of $R(K)$ and $R(K^*)$  can be explained  by considering  NP contributions to the Wilson-coefficients in Eq.  \ref{effective}: $C_i=C^{SM}_i+\Delta C_i$.
In RS bulk custodial models, there are four contributions to FCNC at tree level: $X\in {Z_{SM,Z_X,Z_H,\gamma^{(1)}}}$.
Using Eq. \ref{neutral},  the expression for the coupling of the SM fermions to these NP states is given  as:
\begin{equation}
\mathcal{L}_{NP}\subset \sum_{X=Z_{SM},Z_H,Z_X,\gamma^{(1)}}X_{\mu}\left[\alpha^{bs}_L(X)(\bar s_L\gamma^\mu b_L)+\alpha^{bs}_R(X)(\bar s_R\gamma^\mu b_R)+\bar l\left(\alpha^l_V(X)\gamma^\mu-\alpha^{l}_{A}(X)\gamma^\mu\gamma^5 \right)l \right]
\end{equation}
where $\alpha^l_{V,A}(X)=\frac{\alpha^l_L(X)\pm\alpha^l_R(X)}{2}$ and are defined in Appendix \ref{rsintro}.
Using these expressions, the Wilson co-efficients for each gauge field $X$ can now be written as:
\begin{eqnarray}
\Delta C_9&=&-\frac{\sqrt{2} \pi}{M_X^2 G_F\alpha}\alpha^{bs}_L(X) \alpha^l_{V}(X),\;\;\;\;\;\;\Delta C'_9=-\frac{\sqrt{2} \pi}{M_X^2 G_F\alpha}\alpha^{bs}_R(X) \alpha^l_{V}(X)\nonumber\\\Delta C_{10}&=&\frac{\sqrt{2} \pi}{M_X^2 G_F\alpha}\alpha^{bs}_L(X) \alpha^l_{A}(X),\;\;\;\Delta C'_{10}=\frac{\sqrt{2} \pi}{M_X^2 G_F\alpha}\alpha^{bs}_R(X) \alpha^l_{A}(X)
\label{wc}
\end{eqnarray}
In deriving the above expressions, we assumed that the up-sector quark are in the mass-diagonal basis and $D_{L,R}\sim V_{CKM}$. We now discuss two different possibilities for the fits to the data:\\

1) Scenario \textbf{A}:  This scenario is characterized by the possibility of relatively larger coupling of the lepton singlets to the NP than the lepton doublets. The doublet coupling to NP is assumed to be universal. The scanning range for the doublets and the muon singlets is chosen to be $0.45<c<0.55$. 
Thus even though this scenario can admit a relatively larger coupling of muon singlets to NP than the corresponding doublets, the other possibility still exists.

\begin{figure}
  	\begin{center}
\includegraphics[width=7.2cm,height=9cm,keepaspectratio]{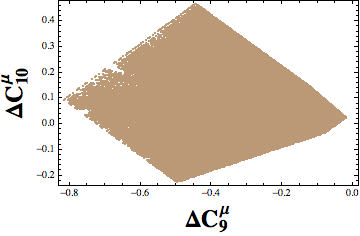} \includegraphics[height=9cm,keepaspectratio,width=7.2cm]{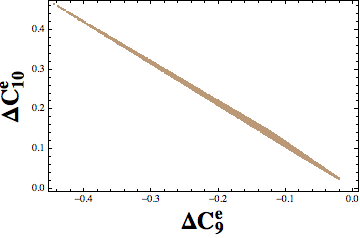}
  			\includegraphics[width=7.2cm,height=9cm,keepaspectratio]{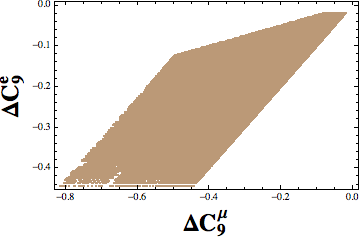} \includegraphics[height=9cm,keepaspectratio,width=7.2cm]{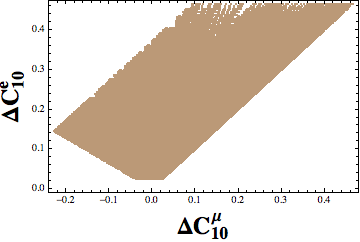}
  	\end{center}
  	\caption{ Scenario \textbf{A}: Plots gives the correlation in the $C_9$ and $C_{10}$ parameter plane for both the electron and the muon. 
  		We use $M_{KK}=3~TeV$}
  	\protect\label{hypothesis1}
  \end{figure}

There are several analyses which point towards the possibility of no NP in the primed operators \cite{Altmannshofer:2017yso,Geng:2017svp}.
 Following this paradigm, we demand that  NP contributions to $\Delta C'_{9,10}$ in this case too must be consistent with zero. One possible way to implement this is by assuming that the right handed down quarks couple similarly to NP. Fig. \ref{overlap} gives the coupling of SM fermions to the gauge KK states as a function of bulk mass parameter $c$. Universality in the down sector couplings can be ensured by choosing
  $c_{d_R,s_R,b_R}>0.55$.  The ranges chosen for $c$ parameter scan is: $c_{Q_3}\in[0.4,0.5]$, $c_{\mu_L}=c_L\in[0.45,0.55]$ and $c_{\mu_R}\in[0.45,0.55]$. The scan in this scenario has some interesting features:
  The $\Delta C^\mu_9-\Delta C^\mu_{10}$ plot has an interesting feature where $\Delta C^\mu_{10}$ may vanish while the contribution to $\Delta C^\mu_9$ is considerable. This is an artifact of the similar range of scan chosen for the lepton doublets and the muon singlets and corresponds to the explicit case where $\alpha^l_L(X)=\alpha^l_R(X)$. Two minor possibilities exist in this scenario:\\
 a) $\Delta C^\mu_{10}$ vanishes and the contributions to $\Delta C^e_{9,10}$ are very small. This essentially reduces to a one-dimensional fit with contributions mainly from $\Delta C^\mu_{9}$. The best fit for this case is $\Delta C^\mu_{9}=-1.5$ with a $2\sigma$ region $[-2.9,-1.73]$ \cite{Altmannshofer:2017yso,Geng:2017svp}. From the bottom left plot of Fig.\ref{hypothesis1}, we find that we can obtain a fit only in the $3\sigma$ region where the electron contribution is negligible.  \\
 b) The tight condition $1-D$ scenario can be relaxed, once the electrons also contribute considerably. This opens up  the possibility of a four dimensional fit for the explanations of lepton flavour non-universality. The results for this case are shown in Fig.\ref{hypothesis1} which gives the correlation between the Wilson co-efficients for both the electron and muon.
 Lower row of Fig. \ref{hypothesis1} gives the different correlations:   $C^\mu_9-C^e_9$ (left) and $C^\mu_{10}-C^e_{10}$ (right). The $2$-$\sigma$ regions for a 4D fit to the data is \cite{Hurth:2016fbr,Mahmoudi}
 \begin{equation}
 C_9^{\mu}/C^{SM}_9\in[-0.33,0.06]\;\;\;C_9^{e}/C^{SM}_9\in[-2.23,0.74]\;\;\;C_{10}^{\mu}/C^{SM}_{10}\in[-0.29,0.14]\;\;\;C_{10}^{e}/C^{SM}_{10}\in[-2.60,0.60]
 \end{equation}
 With a four dimensional case, it is relatively easier to find solutions which satisfy the above regions.
  Thus, this is an explicit realization of a scenario where contribution to the $B$ anomalies are  due to non-universal coupling of the $\mu_R$. However, it must be noted that this is not the only contribution and the doublets also have a  non-negligible contribution.

  
   Further the non-negligible values of the $\Delta C^e$ is due to left doublets having $c\sim 0.5$ thereby resulting a mildly larger coupling to the NP states than would be expected of states having $c\geq 0.55$. 
    Fitting the muon mass for the choices of $c$ used to determine the values of the Wilson-coefficients in Fig.\ref{hypothesis1}  requires choosing the $\mathcal{O}(1)$ Yukawa $\sim 0.03$. As will be seen in Section \ref{MFV}, though slightly fine tuned with regards to the fit to the muon mass, this scenario is more favorable with regards to suppressing FCNC in the lepton sector.\\

2) Scenario \textbf{B}: This is roughly the mirror image of the first scenario where the non-universality is now transferred to the lepton doublets while the singlets are closer to the UV brane and their coupling to the NP is universal. A distinct feature of this scenario is the sign of the $\Delta C_{10}^\mu$, which is exclusively positive as compared to both possibilities obtained earlier. This is mainly due to $\alpha^l_L(X)>\alpha^l_R(X)$ for the leptons. Thus only $c_{\mu_L,\tau_L}<0.5$ while $c_{e_L}>0.55$. Further, without loss of generality we can assume that $c_{\tau_L}<c_{\mu_L}$ resulting in the left handed tau doublets being more composite than the first two generations.
 The singlets for all there generations in the lepton sector satisfy $c>0.55$. These choices result in the contribution to $\Delta C^e_{9,10}$ much smaller  than  $\Delta C^\mu_{9,10}$, with its magnitude being 
 at most $\sim 0.2$. For most of the region, where the value of $C^e_{9,10}$ is an order of magnitude less,  it effectively reduces this to a 2-D fit. 

\begin{figure}
	\begin{center}
		\begin{tabular}{cc}
			\includegraphics[width=7.2cm,height=9cm,keepaspectratio]{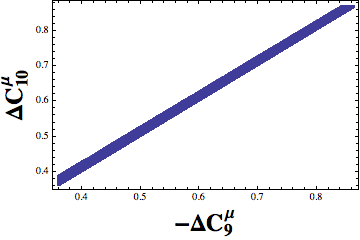}&\includegraphics[width=7.2cm,height=9cm,keepaspectratio]{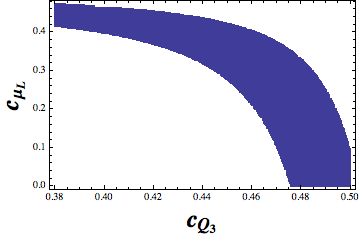}
		\end{tabular}
	\end{center}
	\caption{Scenario \textbf{B}: Left plot gives the distribution for $\Delta C_9$ and $\Delta C_{10}$.  The corresponding $c$ parameters ranges are given in the right plot.  }
	\protect\label{hypothesis2}
\end{figure}

Top left plot of Fig.\ref{hypothesis2} gives the correlation in the $\Delta C^\mu_9$-$\Delta C^\mu_{10}$ plane. 
We accept points which satisfy $0.36<|\Delta C^\mu_{9,10}|<0.87$.
It can be clearly seen that there exist solutions for which $\Delta C^\mu_9=-\Delta C^\mu_{10}$ thereby reducing it to a 1-D fit as discussed in \cite{Capdevila:2017bsm,Altmannshofer:2017yso}.
The only two relevant parameters for the fits to the B-anomalies are $c_{Q_3}-c_{\mu_L}$ and the correlation is shown in right plot of Fig. \ref{hypothesis2}.
This demonstrates a mild  degree of compositeness in one of the parameters or a partial compositeness in the muon doublets and third generation quark doublet is sufficient to explain the anomalies to the data.

 \textbf{Implications of composite leptons:} The two scenarios under consideration involve composite leptons (electrons or muons). It is necessary to address  the possible ramifications on the indirect observables (precision electroweak, $g-2$ \textit{etc.}) as well as constraints from direct $Z'\rightarrow ll$ searches. We being with a discussion on the modification to the $Z\rightarrow ll$ couplings for each of the two scenarios considered above:
 \begin{itemize}
 	\item Scenario A. In this case, the lepton doublets have a universal bulk wave-function with bulk mass parameter $c>0.5$, implying they are elementary. The muon singlets are relatively closer to the IR brane in this case. We choose them to lie in the range $[0.4,0.5]$.
 	We note that like the doublets, the down type singlets are also embedded in custodial representations of $SU(2)_L\times SU(2)_R$. As a result, corrections to the $Z\rightarrow\mu\mu$ couplings tend to cancel between the $Z'$ and $Z_X$ KK states. 
 	
 	\item Scenario B. This is characterized by the muon doublets being relatively closer to the IR brane. However, as noted in the first case, the doublets are  also embedded in a bi-doublet representation of $SU(2)_L\times SU(2)_R$ and hence are also protected against dangerous contributions to $Z\rightarrow\mu\mu$. Further, in light of the constraint on $c_{Q_3}$, the best fit to the neutral current anomaly data also corresponds to the left handed muons having non-zero elementary component, therefore making it fairly safe from corrections to the other precision observables.
 	 \end{itemize}
 	Next possibility is the additional contributions  to the $g-2$ of the muon. In this case, both the gauge boson and Higgs diagrams contribute.
 	 The gauge boson contributions to the $g-2$ of the muon were computed explicitly in the 5D framework and given as \cite{Beneke:2012ie}
 	 \begin{equation}
 	 \Delta a_\mu=8.8\times 10^{-11}\left(\frac{TeV}{\Lambda_{IR}}\right)^2
 	 \end{equation}
 	 where $\Lambda_{IR}=e^{-kr\pi}M_{Pl}$ is the warped down curvature scale. This contribution is independent of the bulk mass and the Yukawa parameters of the model.
 	 For the case where the lowest KK excitations are $\sim 3$ TeV, contributions to $\Delta a_\mu$ is suppressed. Regarding the Higgs contribution, constraints from 
 	 $\mu\rightarrow e\gamma$ make the corresponding contributions to $\Delta a_\mu$ below $10^{-12}$ in the cases as considered here.
 	
\begin{figure}
 	\begin{center}
 		\begin{tabular}{c}
 			\includegraphics[width=7.2cm]{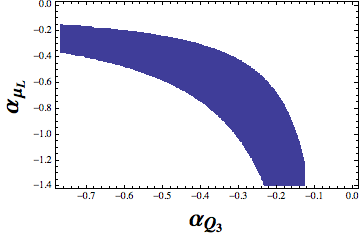}
 		\end{tabular}
 	\end{center}
 	\caption{Copupling of the heavy $Z'$ bosons to the heavy quarks and leptons. }
 	\protect\label{coupling}
 \end{figure}

 	 We now move to a discussion on bounds from direct searches
 For a given $Z'$-mass, bounds exist on the strength of its couplings to the fermions.
 While the coupling to the light quarks ($g_q$) influences its  production cross-section ($\sigma_Z'$), the coupling to the other fermions ($g_{Q_3,t,e,\mu,..}$) places a bound on the corresponding branching fractions.
  We refer to the analysis of \cite{Allanach:2015gkd}, where for $m_{Z'}=3$ TeV  $\mathcal{O}$(1) couplings were allowed even with the approximation $g_{Q_3}=g_\mu=g_q$. Fig.\ref{coupling} gives the distribution of these couplings for Scenario B  as they involved muons with higher degree of compositeness. We note that in this case the coupling of the muons to the $Z'$, $|\alpha_\mu|>1$ only when $\alpha_{Q_3}$ decreases, implying a increasingly elementary third generation quark doublet. Thus, in addition to the PDF suppression of the production cross-sections (for $b$ quarks) it is also accompanied by a small coupling.
  Additionally, coupling of the light quarks to these states $\sim 0.02$. A combination of both these factors will result in the choice of $M_{KK} =3$ TeV consistent with the current bounds.

\section{Kaon decays}
\label{kaon}
In the previous section we discussed   two different possibilities to explain the $B$ anomalies in the same framework. In the event of its confirmation, it is essential to pin down the exact parameter space of the model. This may be possible by correlating the solutions to rare $K$ decays:
$K^+\rightarrow\pi^+\nu\nu$ and $K_L\rightarrow\pi^0\nu\nu$ are likely to constitute the next probe towards the possible existence of NP. They correspond to $s\rightarrow d\nu\nu$ form of transitions and are generally correlated to $b\rightarrow sll$ transitions in most NP scenarios.

 The operators  responsible for  $s\rightarrow d\nu\nu$ form of transitions can be parametrized by the following effective Lagrangian:
\begin{eqnarray}
\mathcal{L}=\frac{4G_F\alpha}{2\sqrt{2}\pi}V^*_{ts}V_{td}C_{ds,l}\left(\bar s_L\gamma_\mu d_L\right)\left(\bar{\nu_l}\gamma^\mu\nu_l\right)
\label{kaoneffective}
\end{eqnarray}
The Wilson coefficient $C_{ds,l}$ in the SM is given as:
\begin{equation}
C^{SM}_{ds,l}=-\frac{1}{s^2_{\theta_w}}\left(X_t+\frac{V^*_{cs}V_{cd}}{V^*_{ts}V_{td}}X^l_c\right)
\label{kaonwc}
\end{equation}
where $X_t$ and $X^l_c$ are the loop functions (Inami-Lim) for the top and charm contribution respectively and given as: $X_t=1.481\pm0.009$ and $\frac{1}{3}\sum_l\frac{X^l_c}{\lambda^4}=0.365\pm0.012$ \cite{Buchalla:1998ba} respectively. Using this,
the branching ratio for $K^+\rightarrow\pi^+\nu\nu$ and $K_L\rightarrow\pi^0\nu\nu$ is given as:
\begin{eqnarray}
\mathcal{B}(K^+\rightarrow\pi^+\nu\nu)&=&\frac{\kappa_+(1+\Delta_{em})}{3}\sum_{l=e,\mu,\tau}\Big |\frac{V^*_{ts}V_{td}}{\lambda^5}X_t+\frac{V^*_{cs}V_{cd}}{\lambda}\left(\frac{X^l_c}{\lambda^4}+\delta P^l_c\right)\Big |^2 \nonumber \\
\mathcal{B}(K_L\rightarrow\pi^0\nu\nu)&=&\frac{\kappa_L}{3}\sum_{l=e,\mu,\tau}\left(\frac{\text{Im}(V^*_{ts}V_{td})}{\lambda^5}X_t\right)^2
\end{eqnarray}
where $\kappa_L=2.231\pm0.013\times 10^{-10}(\lambda/0.225)$, $\kappa_+=5.173\pm0.025\times 10^{-11}(\lambda/0.225)$, $\Delta_{em}=-0.003$ \cite{Mescia:2007kn} and $\delta P^l_{c,u}=0.04\pm0.02$ \cite{Isidori:2005xm}. The individual values of $X^l$ were obtained from Table 1 of \cite{Buchalla:1993wq}: $X^{e,\mu}=11.18\times 10^{-4}, X^{\tau}=7.63\times 10^{-4}$.

We now consider the NP contributions to the process $s\rightarrow d\nu\nu$ given in Eq. \ref{kaoneffective}.  
In the bulk custodial model under consideration, the effective lagrangian for the process is given as
\begin{eqnarray}
\mathcal{L}_{s\rightarrow d\nu\nu}\equiv\left[ \alpha^{sd}_L(\bar s_L\gamma^\mu d_L) + \alpha^{sd}_R(\bar s_R\gamma^\mu d_R) \right](\bar{\nu_l\gamma_\mu\nu_l})\alpha^l_L
\end{eqnarray}
In general this includes both the left handed and the right handed current in the quark sector signaling a possible deviation from the $(V-A)(V-A)$ structure given in Eq. \ref{kaoneffective}. This aspect was explored in great detail in \cite{Blanke:2008yr}.  
We discuss this process in the context of the two scenarios discussed in Section \ref{btos}. It is worth stressing at this point that the $s\rightarrow d\nu\nu$ transitions only depend on the left handed $c_L$ parameters for the leptons, while in the quark sector both $c_{Q_3}$ and $c_{b_R}$ play a role. However, since we assumed only the third generation doublets to have $c_{Q_3}<0.5$, there are no tree-level FCNC in the right handed sector. The contribution can be quantified by making the following change to the $X_t$ in Eq. \ref{kaonwc}:
\begin{equation}
X_t\rightarrow X_t + \sum_{X=Z_{X,H}\gamma^{(1)}}\frac{\sqrt{2}2\pi}{4 G_f\alpha}\frac{\alpha^{sd}(X)\alpha^{l}_L(X)}{M^2_{KK}}
\end{equation}

\begin{figure}
	\begin{center}
		\begin{tabular}{cc}
			\includegraphics[width=7.2cm]{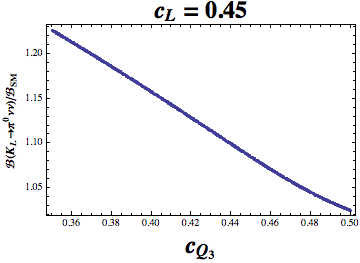}&\includegraphics[width=7.2cm]{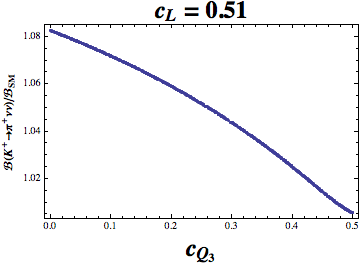}
		\end{tabular}
	\end{center}
	\caption{Scenario \textbf{A}: Plots depicting the excess over the SM expectation for the $K$ decays modes. The $c$ parameters for the doublets is universal and chosen to be $c_L=0.51$. }
	\protect\label{kaons1}
\end{figure}

We consider the following ratio for both the decays $\mathcal{B}^i_{total} / \mathcal{B}^i_{SM}\;\;\;\text{for}~i=K_L,K^+$
and evaluate it for the two scenarios discussed earlier:\\
1) Scenario \textbf{A}: This case is characterized by the universality in the left handed lepton sector. Since  neutrinos in the final state are left handed, only $c_L$ (parameter for the lepton doublets) will play a role in its computation. To stress the fact that $B$ anomalies are explained purely due to non-universality in the right handed sector for leptons we choose : $c_{\mu_R}\sim 0.48$ for the muon singlet while $c_L\sim 0.45$ for all three generations.
Fig. \ref{kaons1} gives plot of $\mathcal{B}^i_{total} / \mathcal{B}^i_{SM}$ computed as a function of $c_{Q_3}$ and evaluated for $c_L=0.51$. This corresponds to the parameter space of the hypothesis under consideration. It can be seen that for both the decays, the ratio is very close to the SM prediction thereby predicting no net enhancement.  \\
\begin{figure}
	\begin{center}
		\begin{tabular}{cc}
			\includegraphics[width=7.2cm]{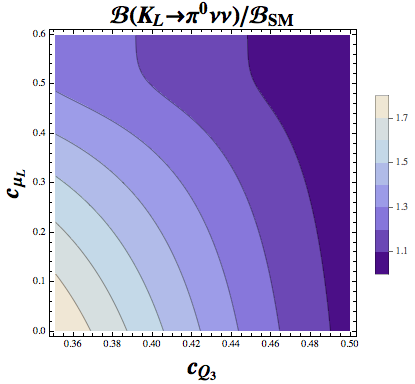}&\includegraphics[width=7.2cm]{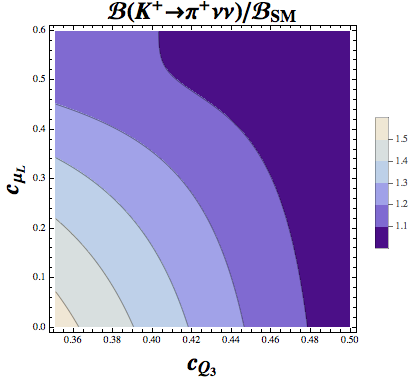}
		\end{tabular}
	\end{center}
	\caption{Scenario \textbf{B}: Plots depicting the excess over the SM expectation for the $K$ decays modes.  $c_{\tau_L}=0.4$ and $c_{e_L}=0.6$ are fixed for the computation while $c_{\mu_L}$ is varied.}
	\protect\label{kaons2}
\end{figure} 

2) Scenario \textbf{B}: This case is characterized by non-universality in the left handed lepton sector while the NP coupling to the right handed singlets are universal.
Fig. \ref{kaons2} gives the ratio $\mathcal{B}^i_{total} / \mathcal{B}^i_{SM}$ for both the kaon decays as a function of $c_{\mu_L}$ and $c_{Q_3}$. We note that for this scenario, the enhancements is not only due to muon doublets being composite, but even the compositeness of the tau doublets $L_3$ aids in this case receiving a larger enhancement relative to first case.  The region consistent with the $b\rightarrow sll$ leads to enhancement of $\sim 1.2-1.6$, depending on the value of $c_{\mu_L}$ and $c_{Q_3}$. This is an useful example where a more accurate measurement of certain process may help in narrowing down the NP parameter space.

\subsection{Correlations between $K^+\rightarrow\pi^+\nu\nu$ and $K_L\rightarrow\pi^0\nu\nu$ }
\begin{figure}[htb!]
	\begin{center}
		\begin{tabular}{cc}
			\includegraphics[width=7.2cm]{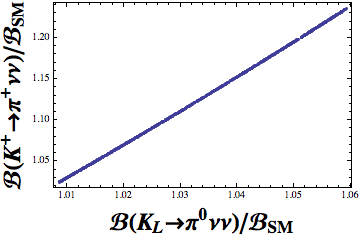}&	\includegraphics[width=7.2cm]{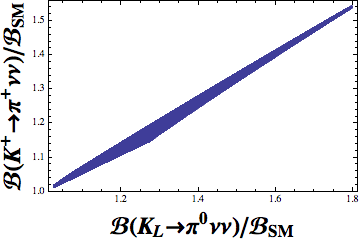}
		\end{tabular}
	\end{center}
	\caption{ Correlation between $K^+\rightarrow\pi^+\nu\bar{\nu}$ and $K_L\rightarrow\pi^0\nu\nu$ for Scenario A (left) and Scenario B (right) }
	\protect\label{correlation}
\end{figure}

In the setup under consideration, both the $K_L\rightarrow\pi^0\nu\nu$ and $K^+\rightarrow\pi^+\nu\nu$ are almost linearly correlated. This can be attributed to the fact that both are described by $s\rightarrow d$ transitions and with a operator structure exactly similar to the SM \textit{viz.} $(V-A)(V-A)$. Fig. \ref{correlation} gives the correlation between the two processes for both the scenarios under consideration.

\section{Charged current processes: $b\rightarrow cl\nu$ }
\label{RDstar}
\begin{figure}
	\begin{center}
		\begin{tabular}{c}
			\includegraphics[width=7.2cm]{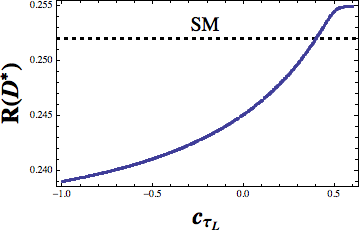}
		\end{tabular}
	\end{center}
	\caption{$R(D^*)$ in custodial RS models as a function of $c_{\tau_L}$. Horizontal line represents the SM value. }
	\protect\label{RD*}
\end{figure}

There has also been a hint of LFU violations in the charged current sector through the measure of the following observable:
\begin{equation}
R(D^*)=\frac{\mathcal{B}(\bar B\rightarrow D^*\tau^-\bar \nu_\tau)}{\mathcal{B}(\bar B\rightarrow D^*l^-\bar \nu_l)}\;\;\text{where}~l=\mu, e
\end{equation}
This corresponds to charged-current transitions in the $b\rightarrow cl\nu$ sector and can be parametrized by the following effective Lagrangian.
\begin{equation}
\mathcal{L}_{b\rightarrow c l\nu}\subset \frac{4 G_f}{\sqrt{2}}V_{cb}\left[C_\tau (\bar c\gamma^\nu P_L b)(\bar \tau \gamma_\nu (U\nu))+C_\mu (\bar c\gamma^\nu P_L b)(\bar \mu \gamma_\nu (U\nu))\right]
\end{equation}
where $C_a=C^{SM}_a+\alpha_a$, $a=(e,\mu),\tau$ and $\alpha_a$ representing the NP contribution. Similar to the neutral currents, the charged currents also receive three contributions:$W_{SM},W_H,W_X$. Using Eq. \ref{chargedcurrent}, the NP contributions to the $b\rightarrow c l\nu$ process is given as \footnote{For simplicity we assume massless neutrinos}:
\begin{eqnarray}
\alpha_{\tau,\mu}&=&\frac{m^2_W}{M^2_{KK}}\left[-\sqrt{2\pi kR}+I(q)+\frac{(V^*_{u_L})_{32}(V_{d_L})_{33}}{V_{cb}}\left(I(b_L)-I(q)\right)\right]I_{\tau,\mu}\nonumber\\
&-&\frac{m^2_W}{M^2_{KK}}\sqrt{2\pi kR}\left[I(q)+\frac{(V^*_{u_L})_{32}(V_{d_L})_{33}}{V_{cb}}\left(I(b_L)-I(q)\right)\right]
\label{charged}
\end{eqnarray}
The volume element $\sqrt{2\pi kR}$ represents the correction to the SM gauge boson coupling
and $I(q)$ is the coupling for the first two generation quarks which is taken to be universal.
Using this, we can express $R(D^*)$ as \cite{Megias:2017ove}:
\begin{equation}
R(D^*)=2R(D^*)\vert_{SM}\frac{\vert 1+ \alpha_\tau\vert^2}{1+\vert 1+ \alpha_\mu\vert^2}
\end{equation}
where it is assumed that NP only couples to the second and third generation leptons. We fix $c_{\mu_L}=0.45$ for the computation.
From Fig \ref{overlap} it is clear that the numerical value of $I$ for the RS geometry is $I\lessapprox 8$, where the maximal value corresponds to brane localized fermions. Even in this extreme case $I(b_L)\simeq\sqrt{2\pi kR}$ resulting in $-\sqrt{2\pi kR}+I(q)+\frac{(V^*_{u_L})_{32}(V_{d_L})_{33}}{V_{cb}}\left(I(b_L)-I(q)\right)\leq 0$. For the case where $c_{\tau_L}\sim 0.4$ the net contribution to $\alpha_\tau$ in  Eq. \ref{charged} is close to zero resulting in the net $R(D^*)$ being consistent with the SM. As the $\tau_L$ becomes increasingly elementary, the net contribution flattens out as shown in Fig. \ref{RD*}.
However, if the $\tau_L$ is pushed closer to the IR brane, $I_\tau$ increases resulting in $\alpha_\tau <0$: thereby   the net contribution being $R(D^*)<R(D^*)\vert_{SM}$ as shown in Fig. \ref{RD*}. Thus  this scenario is highly predictive and subject to validation with future measurements in this sector.

\section{Leptonic MFV}
\label{MFV}
The non-universal coupling of the leptons to the gauge KK states  also give rise  to additional contributions to different
 FCNC processes in leoton sector. 
For KK scales within the reach of LHC, these contributions can be particularly large 
owing to strong upper bounds on processes in the $1$-$2$ generation of leptons: $\mu\rightarrow e \gamma$, $\mu\rightarrow eee$, $\mu-e$ conversion \cite{Agashe:2006iy}. The current experimental bounds on these processes are given in Table \ref{br}.
\begin{table}
	\begin{center}
		\begin{tabular}{ |c | c |  }
			\hline
			Process&Experimental(Upper Bound)\\
			\hline			
			$\mathcal{B}$($\mu\rightarrow e\gamma$) & $4.2\times 10^{-13}$ \cite{TheMEG:2016wtm} \\
			$\mathcal{B}$($\mu\rightarrow eee$) & $1.1\times 10^{-12}$\cite{Bellgardt:1987du} \\
			$\mathcal{B}$($\mu- e$) Conv(Ti) & $6.1\times 10^{-13}$ \cite{Kaulard:1998rb}\\
			\hline  
		\end{tabular}
	\end{center}
	\caption{Experimental upper bound for the branching fraction of leptonic flavour observables in the $1$-$2$ sector.} \label{br}
\end{table}
The large contributions are primarily due to the misalignment between the Yukawa coupling matrix and the bulk mass parameters which determine the nature of the fermionic profiles in the bulk. 
A successful model explaining anomalies in $B$ sector with a relatively low NP scale must also satisfy constraints coming from the non-observation of FCNC processes in the lepton sector. 

Different scenarios have been proposed to alleviate these constraints:
One possibility is to use discrete symmetries as was demonstrated in \cite{Csaki:2008qq}. Another alternative is the implementation of MFV in 5D \cite{Fitzpatrick:2007sa} and will the be focus of attention in this paper. The original MFV ansatz assumed that Yukawas are the only source of flavour violation \cite{DAmbrosio:2002vsn}. It was extended to 5D warped scenarios \cite{Fitzpatrick:2007sa} and  studied in  \cite{Fitzpatrick:2007sa,Chen:2008qg,Iyer:2012db} for the leptonic sector, This 
   can be understood as follows: As stated earlier, the large contributions to FCNC are due to the misalignment between the flavour violating contribution $YY^\dagger$ and the mass-squared matrix $m^2\sim (F(c)YY^\dagger F(c))$ where $F(c)=Diag(f(c_e),f(c_\mu),f(c_\tau))$. Since $Y$ and $F(c)$ in general do not commute, diagonalization of $m^2$ does not imply diagonalization of $YY^\dagger$. 
  This misalignment can however be reduced if the bulk mass parameters are written in terms of the 5D Yukawa parameters   as:
\begin{eqnarray}
c_L=a_L I+b_LY_EY_E^\dagger +d_L Y_NY_N^\dagger\;\;\;c_E=a_E I+b_E Y_E^\dagger Y_E \;\;\;c_N=a_N+b_N Y_N^\dagger Y_N
\label{align}
\end{eqnarray} 
where $a_i,b_i\in \Re$.
Furthermore we assume the presence of a bulk flavour symmetry group in the leptonic sector is given $SU(3)_L\times SU(3)_E\times SU(3)_N$. Using the flavour symmetry we choose  a basis in which $Y_E$ is diagonal. Thus flavour violations are encoded in $Y_N$ which  transforms under the flavour group as $Y_N\rightarrow V_{5}Y_N$ where $V_5$ is the anarchic mixing matrix. Without loss of generality we can choose $V_5=V_{PMNS}$. Simultaneously we also assume that the corresponding bulk flavour symmetry in the hadronic sector is broken with the corresponding $c$ parameters not conforming to similar relations like Eq. \ref{align} in the hadronic sector. 
With this background now discuss its implementation for the two scenarios discussed in Section \ref{btos}.\\
a) Scenario \textbf{A}: In this case the lepton doublets are localized closer to the UV brane and NP effects are due to the coupling of the right handed singlets.
We assume $c_L\simeq 0.51$ (From   Fig. \ref{hypothesis1}) for all three generations, similar to  \cite{Perez:2008ee,Chen:2008qg,Iyer:2012db}. The only difference in this case is that the second and third generation charged singlets must be localized closer to the IR brane to satisfy the $B$ anomalies. In view of this the $c$ values for the charged lepton singlets are chosen as $c_E=\{0.4,0.48,0.764\}$ and corresponding Yukawa couplings are $Y_E=\{0.77,0.03,0.32\}$. The $c_E$ are written in terms of $Y_E$ by choosing $a_E=0.49,b_E=-0.84$ in Eq. \ref{align}. For $c_L$ we choose $a_L\gg b_L,d_L$ to preserve the universality for the left handed doublets.

With regards to fits for the neutrino masses we make choices  for $c_N$ and $Y_N$ as: 
  $c_N=\{1.17,1.17,1.21\}$ and $Diag(Y_N)\equiv\{0.1,0.1,0.134\}$ leading to the following fit for the neutrino oscillation data:
  \begin{equation}
  \Delta m^2_{sol}=7.7\times 10^{-5}~eV^2\;\;\;\Delta m^2_{atm}=1.98\times 10^{-3}~eV^2
  \end{equation}
  corresponding to an inverted hierarchy spectrum. It is to be stressed that these values are not a prediction of the owing the large parameter space as shown on \cite{Iyer:2012db}.
  
  Corresponding to these choices, the $\mathcal{B}$($\mu\rightarrow e\gamma$) is given as \cite{Perez:2008ee}:
\begin{equation}
\mathcal{B}(\mu\rightarrow e\gamma)=4\times 10^{-8}\times (Y_NY_N^\dagger)^2_{12}\frac{3 TeV}{M_{KK}}
\end{equation}
 where
\begin{equation}
Y_N=\begin{bmatrix}
0.0823679&0.0548227& 0.0194184\\ -0.0477093& 0.0531867& 0.0937521\\ 0.0306488& -0.0645418& 0.0937521
\end{bmatrix}
\end{equation} 
leading to $\mathcal{B}(\mu\rightarrow e\gamma)\sim 2.5\times 10^{-14}$. Similarly the other $1-2$ transitions are also within the experimental bounds quoted in Table \ref{br} for $M_{KK}=3$ TeV.  \\
b) Scenario \textbf{B}: In the earlier case we saw that the universality of the lepton doublet parameters $c_L$ was critical in the implementation of the MFV ansatz and also satisfactory fit to the neutrino mases. Universality ensures that the values of the rotation matrix $U_{ij}\sim\frac{f_{L_i}}{f_{L_i}}\sim \mathcal{O}(1)$ corresponding to the elements of the PMNS rotation matrix.
In this case however, the lepton doublets are not universal. As a result fits to the neutrino data also require hierarchical choices in $c_N$. This makes the implementation of $c_N$ proportional to $Y_N^\dagger Y_N$ in \ref{align} extremely challenging and difficult to achieve, possibly requiring a more complicated parameter scan which is outside the scope of this exercise.

 \section{Conclusions} 
 Custodial RS models offer an interesting avenue to explore flavour physics. An interesting feature of the model is that while $R(K),R(K^*)$ can be easily accommodated, the charged current anomalies are consistent with the SM. We fit the anomalies in the $b\rightarrow sll$ transitions for two different scenarios:\\ \textbf{A}) Right handed leptons have a tendency to be more composite than the left handed leptons. This case is characterized by small but non-negligible contributions to $\Delta C^e_{9,10} $ and \textbf{B}) Left handed lepton sector is more composite than the right handed leptons. The latter can mimic a one-dimensional fit with $\Delta C^\mu_{9}=-\Delta C^\mu_{10}<0$.
  We offer a way to distinguish these scenarios by considering $K_L\rightarrow\pi^0\nu\nu$ and $K^+\rightarrow\pi^+\nu\nu$. For the parameter space which fits the anomalies, the former is characterized by their consistency with the SM while the latter can lead to an enhancement of $1.2$-$1.6$. Further, implementation of MFV, necessary to suppress FCNC in lepton sector, is explicitly discussed for Scenario \textbf{A} thereby making it more viable.  An important observation in these scenarios is the presence of non-zero Wilson-coefficients for the electron $C^e_{9,10}$ in both the scenarios discussed above. This is likely to contribute to high precision measurements like the atomic parity violation in Cesium and presents a future direction for this and other models.
  Rare decays and direct CP violation in the Kaon sector \cite{DAmbrosio:2017klp,Chobanova:2017rkj} also present a useful candle to probe NP effects. In the context of the model considered to explain the $B$ anomalies they may offer useful hints on the underlying origin of the non-universality.

\paragraph{Acknowledgements}
GD and AI were supported in part by MIUR under Project No. 2015P5SBHT and by the INFN research initiative ENP. 
We thank A. Buras for a careful reading of the manuscript and many useful comments.
We thank F. Mahmoudi for providing details of the 4D parameter fit. We are grateful to D. Bhatia and A. Kushwaha for careful reading of manuscript. GD would like to thank ICTS/Prog-candark/2017/06 for hospitality.
\appendix
\section{Bulk fields in Custodial RS models}
\label{rsintro}
In this section we review the basic ingredients of the model required to compute the coefficients in Eq. \ref{effective}.
The RS model, with the fermion and gauge fields in the bulk, is beset by large contributions to the $T$ parameters as well as the $Z\bar b_Lb_L$ coupling. This is due to the mixing between the zero-modes and gauge KK-modes induced by electro-weak symmetry breaking \cite{Csaki:2002gy}. One alternative is to consider soft-wall models where this mixing can be considerably weakened \cite{Cabrer:2010si,Cabrer:2011fb}. We consider the other alternative where the bulk gauge symmetry is extended as  \cite{Agashe:2003zs}:
\begin{equation}
SU(2)_L\times SU(2)_R\times U(1)_{B-L}\times P_{LR}
\end{equation}
The Higgs, which transforms as (2,2) under the bulk symmetry group, is localized on the IR brane with the following canonically normalized lagrangian\footnote{The Higgs field is redefined as $H\rightarrow e^{kR\pi}H$ absorb the exponential factors from $\sqrt{-g_{IR}}=e^{-4kR\pi}$ to canonically normalize the kinetic term.}:
\begin{equation}
\mathcal{L}_{Higgs}=(\mathcal{D}H)^2-\mu^2|H|^2+\frac{\lambda}{4}|H|^4
\end{equation}
where $\mathcal{D}=\partial-i\left(g_5W^a\tau_{aL}+\tilde g_5\tilde W^a\tilde\tau_{aR}+\frac{\tilde g_5'}{2}(B-L)\tilde B\right)$ and the lorentz indices have been suppressed for convenience. Thus in addition to the SM gauge boson ($W^\pm_{SM},Z_{SM}$), there are two additional first KK states in both the neutral and the charge current sector \footnote{We consider the  effect of only  the first KK state on the flavour observables. The contributions of the higher KK levels are subleading.}:
$(W^\pm_H,Z_H)$ and $W^{\pm}_X,Z_X$ having similar masses $\sim M_{KK}$. Owing to the mixing induced by the EWSB, these mass eigenstates can be defined in terms of the gauge eigenstates as:
\begin{eqnarray}
Z_{SM}&=&Z^{(0)}-\frac{M_Z^2}{M^2_{KK}}\left(-\sqrt{2kR\pi}(Z^{(1)})+\sqrt{2kR\pi}\cos\phi\cos\psi~Z^{'}\right)\nonumber\\
Z_{H}&=&\cos\zeta~(Z^{(1)})+\sin\zeta~(Z^{'}) \nonumber\\
Z_{X}&=&-\sin\zeta~(Z^{(1)})+\cos\zeta~(Z^{'}) 
\end{eqnarray}
Here $Z^{(0)},Z^{(1)}$ are the zeroth and first KK excitation of the SM gauge eigenstate field $Z$, while $Z'$ is massive field with (approximately) $(-+)$ boundary conditions and defined as: $Z'=\frac{\tilde g_5\tilde W^3-\tilde g'_5\tilde B}{\sqrt{\tilde g_5^2+\tilde g_5^{'2}}}$ 
where $\sin^2\psi\simeq \sin^2\theta_W$ and $\cos\psi={1 \over \sqrt{1+\sin^2\phi}}$. $\zeta$ is the $Z^{(1)}-Z'$ mixing angle. The mass eigenstates for the charged fields $(W^\pm_{SM,X,H})$ can similarly be written in terms of the gauge eigenstates $(W^{(0)\pm},W^{(1)\pm},W^{'\pm})$ as:
\begin{eqnarray}
W^\pm_{SM}&=&W^{(0)\pm}-\frac{m^2_W}{M^2_{KK}}\sqrt{2kR\pi}W^{(1)\pm}\nonumber\\
W^\pm_{H}&=&\cos\chi~(W^{\pm(1)})+\sin\chi~(W^{\pm'}) \nonumber\nonumber\\
W^\pm_{X}&=&-\sin\chi~(W^{\pm(1)})+\cos\chi~(W^{\pm'}) 
\end{eqnarray}
where $\chi$ is the $W^{\pm(1)}-W^{\pm'}$ mixing angle.

While the zero mode gauge fields $X^{(0)}$ where $X=Z,W^\pm$ have a flat profile in the bulk, the higher KK models are characterized by a profile which is peaked near the IR brane. The profiles satisfy the following  differential equation \cite{Davoudiasl:1999tf,Pomarol:1999ad,Gherghetta:2000qt,Huber:2003tu}:
\begin{equation}
[z_n^2\frac{d^2}{dz_n^2}+z_n\frac{d}{dz_n}+(z_n^2-1)](e^{-\sigma}f^{(n)}(y))=0
\end{equation}
where $z_n=\frac{m_n}{k}e^{\sigma}$ and $\sigma =k|y|$.
the solutions to which are as given as
\begin{equation}
f^{(n)}(y)=e^{\sigma}/N_n[J_1(z_n)+b_nY_1(z_n)]
\label{gaugekk}
\end{equation}
where $b_n$ is determined by boundary conditions. Since the boundary conditions for $(Z,W^\pm)^{(1)}$ (++) is different from $(Z,W^\pm)^{'}$, (-,+), the corresponding value of $b_1$ will be different.
The KK photon will have a similar bulk profile but different boundary condition as there are no mass term induced on the IR brane due to EWSB.
The KK masses used in the analysis will correspond to $M_{KK}=3$ TeV.

\textbf{Fermions in the bulk:} In addition to the gauge fields, we also consider bulk fermions as it offers a natural understanding of the Yukawa hierarchy problem:
Since the  fermions in odd-dimensions are vector-like, a bulk 5D   Dirac spinor can be decomposed as:
\begin{equation}
\Psi(x,y)=\Psi_L(x,y)+\Psi_R(x,y)
\end{equation}
with $\Psi_{L,R}=\pm\gamma^5\Psi_{L,R}$, implying that $\Psi_R$ is odd under $Z_2$. Thus only the $\Psi_L\equiv \psi_L(x)f_L(y)$ will have zero mode profile $f^{(0)}_L$. Assuming a bulk mass term parametrized as $m_{\Psi}=c\sigma^{'}$ the extra dimensional profiles can be obtained by solving following coupled equations \cite{Gherghetta:2000qt}:
\begin{eqnarray}
-e^{-\sigma}(\partial_y+c\sigma^{'})f_L^{(n)}(y)=m_nf_R^{(n)}(y) \\
e^{-\sigma}(\partial_y-c\sigma^{'})f_R^{(n)}(y)=m_nf_L^{(n)}(y)
\label{rsfermioneigenvalue}
\end{eqnarray}

The equations decouple for $m_n=0$ and  the zero mode solutions are given as
\begin{equation}
f^{(0)}_{L}=N^{(0)}_{L}e^{(0.5-c)ky}
\label{zeromode}
\end{equation}
where the normalization factors $N^{(0)}_{L}=\sqrt{\frac{k(1-2c)}{e^{(0.5-c)kR\pi}-1}}$ 
From the above equation it is clear that the fermion zero modes are localized towards the UV(IR) for $c>0.5(c<0.5)$. The $c$ parameters play an important role in determing the effective $4D$ Yukawa coupling as:
\begin{equation}
Y^{(4)}=\frac{Y^{(5)}}{k}N^{(0)}_{L}N^{(0)}_{R}e^{(1-c_L-c_R)kR\pi}
\end{equation}
where $Y^{(5)}$ is typically $\mathcal{O}$(1).
The entire spectrum of fermion masses and mixing (lepton and hadron) can be fit by assuming $c$ parameters in the range $-1.5\leq c \leq 1.5$. For our analysis, we assume the fermion doublets to transforms as (2,2) under the gauge group while the singlets $(1,3)$. The coupling constant for fermions with a given representation to the different gauge bosons discussed above. have been outlined in \cite{Blanke:2008yr}. We now discuss the origin of non-universality in bulk RS models.

\subsection{Tree level decays} 
\label{nonuniversality}
As discussed above, the different fermionic generations are localized at different points in the bulk to facilitate a solution to the Yukawa hierarchy problem. While their coupling to 
$Z^{(0)}$ is universal, their coupling to ${Z}^{(1)},Z'$ ( whose profile is peaked near the IR brane) is generation dependent . This coupling depends on the localization of the fermions along the extra-dimension thus giving rise to non-universality. Let $\eta^T =$ 
\{$f^{(1)}_M$,$f^{(2)}_M$, $f^{(3)}_M$\}             
be vector of fermions in the mass basis. Let $a_{ij}^{(1)}$ be a $3\times 3$ matrix denoting the coupling of
SM fermions in the mass basis to a generic KK gauge boson say ${X}^{(1)}$. It is given as
\begin{equation}
a^{ij}_{L,R}=\tilde g~ \eta^T_{L,R} D_{L,R}^\dagger\begin{bmatrix}
I_{f_1}&0&0\\0&I_{f_2}&0\\0&0&I_{f_3} 
\end{bmatrix} D_{L,R} \eta_{L,R}
\label{couplingmatrix}
\end{equation}
where $\tilde g$ is the coupling constant depending on the gauge field and particular representation of the fermion and are given in Appendix in \cite{Blanke:2008yr}. $D_{L,R}$ are $3\times 3$ unitary matrices for rotating the zero mode (SM) fermions from the flavour basis to the mass basis. \textit{I} is the overlap of the profiles of two zero mode fermions and first KK gauge boson and is given by
\begin{equation}
I(c)=\frac{1}{\pi R}\int_0^{\pi R}dy e^{\sigma(y)}(f_i^{(0)}(y,c))^2\xi^{(1)}(y)_{Z^{(1)},Z'}
\label{couplingkkf}
\end{equation}
The off diagonal elements of $a_{ij}^{(1)}$ represent the flavour
violating couplings.
They are given as:
\begin{eqnarray}
a^{{12}}=\tilde g\left(D^*_{21}D_{22}(I(2)-I(1))+D^*_{31}D_{32}(I(3)-I(1))\right)\nonumber\\
a^{{23}}=\tilde g\left(D^*_{12}D_{13}(I(1)-I(2))+D^*_{32}D_{33}(I(3)-I(2))\right)\nonumber\\
a^{{13}}=\tilde g\left(D^*_{21}D_{23}(I(2)-I(1))+D^*_{31}D_{33}(I(3)-I(1))\right)
\end{eqnarray}

\begin{figure}
	\begin{center}
		\begin{tabular}{c}
			\includegraphics[width=7.2cm]{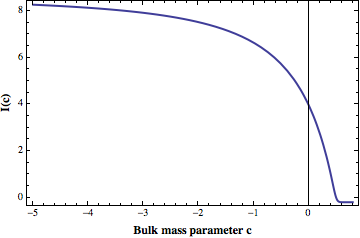}
		\end{tabular}
	\end{center}
	\caption{ Overlap integral $I$ as a function of bulk mass parameter $c$ }
	\protect\label{overlap}
\end{figure}

Fig. \ref{overlap} gives the plot of $I$ as a function of $c$. The integral is universal $I\sim 0.2$ for $c\geq 0.5$. Since the Higgs is localized near the IR brane, $c$ values for all the quark fields with the exception of the third generation will be chosen to be $c>0.5$.
$\xi^{(1)}(y)$ denotes the profile of the first KK gauge boson: $Z^{(1)}$ correesponds to the first KK state of the SM $Z$ with $(+,+)$ boundary condition while $Z'$ is the neutral $SU(2)_R\times U(1)_{B-L}$ with $(-,+)$ boundary condition. As discussed in Section \ref{rsintro}, the breaking of the electroweak symmetry at the IR brane mixes the zero mode gauge boson with the higher modes. 
In the mass basis, the flavour violating couplings is given as:
\begin{eqnarray}
\alpha^{ij}_{L,R}(Z_{SM})&=&\frac{M_Z^2}{M^2_{KK}}\left(-\sqrt{2kR\pi}a^{ij_{L,R}}(Z^{(1)})+\sqrt{2kR\pi}\cos\phi\cos\psi a^{ij_{L,R}}(Z^{'}) \right)\nonumber\\
\alpha^{ij}_{L,R}(Z_{H})&=&\cos\zeta~a^{ij_{L,R}}(Z^{(1)})+\sin\zeta~a^{ij_{L,R}}(Z^{'}) \nonumber\\
\alpha^{ij}_{L,R}(Z_{X})&=&-\sin\zeta~a^{ij_{L,R}}(Z^{(1)})+\cos\zeta~a^{ij_{L,R}}(Z^{'}) 
\label{neutral}
\end{eqnarray}
where $\sin^2\psi\simeq \sin^2\theta_W$ and $\cos\psi=\frac{1}{\sqrt{1+\sin^2\phi}}$. $\zeta$ is the $Z^{(1)}-Z'$ mixing angle. For the computation in neutral current transitions we choose: $\cos\zeta=0.54$ and $\sin\zeta=0.84$.

Similar to  $Z_{H,X}$ the KK photon also contributes to the FCNC with structure similar to Eq. \ref{couplingmatrix} with the replacement that $g\rightarrow eQ$ where $Q$ is the electromagnetic charge and $e=g\sin\theta_W$

Along the same line, the coupling to the charged gauge bosons are given as:
\begin{equation}
W_{SM}:\frac{-ig}{\sqrt{2}}\left(1-\frac{m_W^2}{M^2_{KK}}\sqrt{2\pi kR}I(c_f)\right);\;\;\;\;W_H:\frac{-ig}{\sqrt{2}}\cos\zeta' I(c_f);\;\;\;\;W_X:\frac{ig}{\sqrt{2}}\sin\zeta' I(c_f)
\label{chargedcurrent}
\end{equation} 
Thus similar to the neutral current sector, the mixing between the zero mode and KK mode states induces corrections to the coupling of the fermions to the SM gague bosons. 

\bibliography{biblio1}
\bibliographystyle{JHEP}

\end{document}